\documentclass[11pt]{article}

\usepackage[letterpaper, margin=1in]{geometry}

\usepackage{amssymb}
\usepackage{amsmath}
\usepackage{textcomp}

\usepackage{graphicx}
\usepackage{caption}
\usepackage{subcaption}
\usepackage{booktabs}
\usepackage{array}

\usepackage[affil-it]{authblk}

\usepackage[numbers,sort&compress]{natbib}

\usepackage{hyperref}
\hypersetup{
    colorlinks=true,
    linkcolor=blue,
    citecolor=blue,
    urlcolor=blue
}

\begin{document}

\title{Robustness Analysis of Machine Learning Models for IoT\\Intrusion Detection Under Data Poisoning Attacks}

\author[1]{Fortunatus Aabangbio Wulnye\thanks{Corresponding author: \texttt{fortunatuswulnye@outlook.com}}}
\author[1]{Justice Owusu Agyemang\thanks{\texttt{joagyemang@knust.edu.gh}}}
\author[1]{Kwame Opuni-Boachie Obour Agyekum\thanks{\texttt{kooagyekum@knust.edu.gh}}}
\author[1]{Kwame Agyeman-Prempeh Agyekum\thanks{\texttt{kapagyekum.coe@knust.edu.gh}}}
\author[1]{Kingsford Sarkodie Obeng Kwakye\thanks{\texttt{ksobengkwakye@knust.edu.gh}}}
\author[1]{Francisca Adomaa Acheampong\thanks{\texttt{faacheampong@knust.edu.gh}}}

\affil[1]{Department of Telecommunication, Kwame Nkrumah University of Science and Technology (KNUST), Kumasi, Ghana}

\date{}

\maketitle

\begin{abstract}
Ensuring the reliability of machine learning--based intrusion detection systems remains a critical challenge in Internet of Things (IoT) environments, particularly as data poisoning attacks increasingly threaten the integrity of model training pipelines. This study evaluates the susceptibility of four widely used classifiers, Random Forest, Gradient Boosting Machine, Logistic Regression, and Deep Neural Network models, against multiple poisoning strategies using three real-world IoT datasets. Results show that while ensemble-based models exhibit comparatively stable performance, Logistic Regression and Deep Neural Networks suffer degradation of up to 40\% under label manipulation and outlier-based attacks. Such disruptions significantly distort decision boundaries, reduce detection fidelity, and undermine deployment readiness. The findings highlight the need for adversarially robust training, continuous anomaly monitoring, and feature-level validation within operational Network Intrusion Detection Systems. The study also emphasizes the importance of integrating resilience testing into regulatory and compliance frameworks for AI-driven IoT security. Overall, this work provides an empirical foundation for developing more resilient intrusion detection pipelines and informs future research on adaptive, attack-aware models capable of maintaining reliability under adversarial IoT conditions.
\end{abstract}

\noindent\textbf{Keywords:} Adversarial Machine Learning; IoT Security; Data Poisoning Attacks; Network Intrusion Detection; Threat Intelligence; Cyber-Physical Systems


\section{Introduction}
\label{sec:introduction}
\sloppy
Protecting the integrity of machine learning pipelines has become a central challenge for contemporary IoT security architectures. As Network Intrusion Detection Systems (NIDS) increasingly depend on data-driven inference, the reliability of underlying training data becomes a critical vulnerability. In heterogeneous IoT environments, characterized by lightweight protocols, constrained firmware, and diverse communication behaviors, small perturbations in training samples can influence model generalization in ways that conventional validation practices fail to detect. 

Data poisoning attacks exploit this weakness by injecting manipulated samples or altering feature distributions to reshape decision boundaries, weaken threat detection performance, or embed persistent misclassification patterns within deployed systems \citep{Biggio2012PoisoningSVM, Yuan2019AdversarialExamples, Dunn2020PoisoningRobustness, BaracaldoPoisoningIoTDetection}. These concerns are amplified by forecasts predicting cybercrime damages approaching \$10.5 trillion by 2025 \citep{Morgan2021Cybercrime}, underscoring the urgency of strengthening IoT resilience.
Recent surveys highlight that deep learning–based NIDS have become central to modern IoT defense strategies due to their ability to model complex, evolving traffic behaviours \citep{Farhan2022Survey, al2025comprehensive, santhosh2023comprehensive, ashraf2022iot}. Their adoption has been enabled by large benchmark datasets such as CICIoT2023 \citep{Carlos2023CICIoT2023}, Edge-IIoTset \citep{Ferrag2022EdgeIIoTset}, N-BaIoT \citep{Abbasi2021NBAIoTAnomaly}, and multi-step IIoT datasets \citep{Almseidin2022CyberMultiStep}. These datasets capture diverse threat landscapes spanning botnets \citep{Nazir2023IoTBotnetReview}, sequential IoT attack behaviours \citep{Soe2020IoTBotnetSequential}, and large-scale distributed anomalies \citep{Najafimehr2023DDoSSurvey}. However, poisoning resilience across these datasets remains underexplored, despite growing evidence of poisoning attacks in both centralised and federated architectures \citep{Chen2020FederatedIDS, Li2021DeepFed, Zhang2021PoisonGAN, Chiba2020PoisonDefense}. 

Recent work also shows that clean-label poisoning, where malicious samples appear benign, poses a serious emerging threat to deep neural models \citep{shafahi2018poison}. Motivated by these gaps, this study presents a systematic assessment of how poisoning strategies influence widely used ML models in IoT intrusion detection, including Logistic Regression \citep{Abbasi2021NBAIoTAnomaly}, Random Forest \citep{AnalyticsVidhya2021RandomForest}, Gradient Boosting Machines \citep{Sarcevic2022CybersecurityXAI}, and Deep Neural Networks \citep{DashIoTBiomedicalReview}. By analysing how different attack types perturb model learning, the study clarifies comparative fragility across model families and outlines principles for designing more resilient intrusion detection mechanisms. The work further contributes a unified exploratory analysis of three major IoT datasets, a controlled evaluation of four representative poisoning techniques, and the identification of model-specific behavioural patterns that inform hybrid, poisoning-resilient NIDS architectures for future deployments.

\section{Related Work} 
\label{sec:relatedwork}
\sloppy
Research into securing IoT networks has expanded considerably, particularly in understanding how NIDS perform under increasingly sophisticated cyber threats. Early work established the limitations of signature-based detection, which struggles to cope with dynamic IoT environments and evolving attack strategies \citep{Webster2000DARPAIDS}. This limitation has driven a shift toward anomaly-based and ML-driven approaches capable of modelling behavioural deviations that static rule systems often overlook \citep{Yaokumah2021IoTNIDS, HodoANNIDSThreatAnalysis}. The proliferation of IoT-specific vulnerabilities, stemming from heterogeneous device architectures, lightweight communication protocols, and constrained hardware, further underscores the need for adaptable detection mechanisms \citep{Atzori2010IoTSurvey, Chen2019EdgeDLReview}. These weaknesses have enabled widespread botnet campaigns such as Mirai and Bashlite, reinforcing the critical role of NIDS in securing IoT ecosystems \citep{Yaokumah2021IoTMalwareDetection, Nazir2023IoTBotnetReview}.

ML-based intrusion detection has demonstrated clear advantages in modelling high-dimensional IoT traffic and capturing subtle threat indicators. Ensemble models, such as Random Forests (RF) and Gradient Boosting Machines (GBM), exhibit strong performance due to their robustness to noise and ability to generalize across diverse datasets \citep{AnalyticsVidhya2021RandomForest, Chen2020FederatedIDS}. Deep Neural Networks (DNNs) extend these capabilities by learning hierarchical traffic representations \citep{DashIoTBiomedicalReview}, while Logistic Regression (LR) remains a lightweight option for resource-constrained deployments despite its vulnerability to nonlinear attack patterns \citep{Abbasi2021NBAIoTAnomaly}.

Comprehensive reviews emphasize the growing reliance on ML and DL approaches in IoT intrusion detection, highlighting both their strengths and weaknesses \citep{kikissagbe2024machine, heidari2023internet}. The availability of datasets such as CICIoT2023 \citep{Carlos2023CICIoT2023}, Edge-IIoTset \citep{Ferrag2022EdgeIIoTset}, and N-BaIoT \citep{Abbasi2021NBAIoTAnomaly} has further enabled realistic model benchmarking.

\begin{table*}[t!]
\centering
\scriptsize
\caption{Comparison of Related Works in IoT Security, Intrusion Detection, and Adversarial Machine Learning}
\label{tab2}
\renewcommand{\arraystretch}{1.22}
\setlength{\tabcolsep}{4.2pt} 

\scalebox{0.75}{
\begin{tabular}{l|p{5.6cm}|p{1.05cm}|p{1.20cm}|p{1.20cm}|p{1.55cm}|p{1.10cm}}
\hline\hline
\textbf{Reference} 
& \textbf{Main Contribution} 
& \textbf{IoT Dataset} 
& \textbf{Intrusion Det.} 
& \textbf{Poisoning} 
& \textbf{Defense / Robustness} 
& \textbf{ML Analysis} \\
\hline\hline

\cite{Almseidin2022CyberMultiStep} 
& Multi-step IoT cyberattack dataset enabling realistic model benchmarking. 
& \checkmark & \checkmark & \texttimes & \texttimes & \checkmark \\
\hline

\cite{Biggio2012PoisoningSVM} 
& First gradient-based poisoning attack on SVMs; cornerstone for adversarial ML. 
& \texttimes & \checkmark & \checkmark & \texttimes & \checkmark \\
\hline

\cite{Sugi2020MLTechniquesIoTIDS} 
& Evaluation of ML intrusion detection for IoT networks; strong LSTM performance. 
& \checkmark & \checkmark & \texttimes & \texttimes & \checkmark \\
\hline

\cite{Yaokumah2021IoTNIDS} 
& Survey of IoT IDS frameworks focusing on datasets and ML limitations. 
& \checkmark & \checkmark & \texttimes & \texttimes & \checkmark \\
\hline

\cite{Chiba2020PoisonDefense} 
& Poisoning defense for IoT ML using controlled adversarial sampling. 
& \checkmark & \checkmark & \checkmark & \checkmark & \checkmark \\
\hline

\cite{Dunn2020PoisoningRobustness} 
& Robustness evaluation of ML models under poisoning in IoT environments. 
& \checkmark & \checkmark & \checkmark & \checkmark & \checkmark \\
\hline

\cite{Ferrag2022EdgeIIoTset} 
& Large IIoT dataset enabling advanced IDS benchmarking. 
& \checkmark & \checkmark & \texttimes & \texttimes & \checkmark \\
\hline

\cite{Zhang2021PoisonGAN} 
& PoisonGAN: generative poisoning attacks for federated IoT systems. 
& \texttimes & \checkmark & \checkmark & \texttimes & \checkmark \\
\hline

\textbf{Our Work} 
& \textbf{First unified robustness evaluation of four poisoning techniques across three IoT datasets using multiple ML classifiers, providing comparative insights and defense implications.} 
& \checkmark & \checkmark & \checkmark & \checkmark & \checkmark \\
\hline\hline

\end{tabular}
}
\end{table*}

Parallel to these developments, adversarial machine learning has exposed new attack surfaces in ML-driven NIDS. Foundational work demonstrated how subtle training-time perturbations can corrupt SVMs \citep{Biggio2012PoisoningSVM}, while later studies extended these attacks to deep learning architectures \citep{Yuan2019AdversarialExamples}. Poisoning attacks, including label flipping, synthetic outliers, feature impersonation, and clean-label variants, can distort classification boundaries, embed persistent backdoors, or significantly degrade performance even at low contamination rates \citep{Dunn2020PoisoningRobustness, Chiba2020PoisonDefense, Zhang2021PoisonGAN, Muthukrishnan2017OutlierSurvey}. Clean-label attacks, in particular, pose an emerging threat because they evade standard data sanitization checks \citep{shafahi2018poison}. In distributed settings, federated learning is especially vulnerable due to the potential for compromised clients to propagate adversarial updates \citep{Li2021DeepFed, Chen2020FederatedIDS}.
\

Despite substantial progress, systematic evaluations of poisoning resilience in IoT-focused NIDS remain limited. Existing studies primarily assess models under clean conditions, offering a fragmented understanding of how poisoning affects different ML families or how dataset characteristics influence vulnerability. Table~\ref{tab2} summarises representative work across intrusion detection, adversarial ML, and IoT security. Building on these foundations, the present study provides the first unified, robust evaluation of LR, RF, GBM, and DNN models against four poisoning techniques across three major IoT datasets, delivering comparative insights and highlighting implications for the design of adversarially resilient NIDS.

\section{Experimental Framework}
\label{sec3}
\begin{figure*}[!t]
    \centering
    \includegraphics[width=\textwidth]{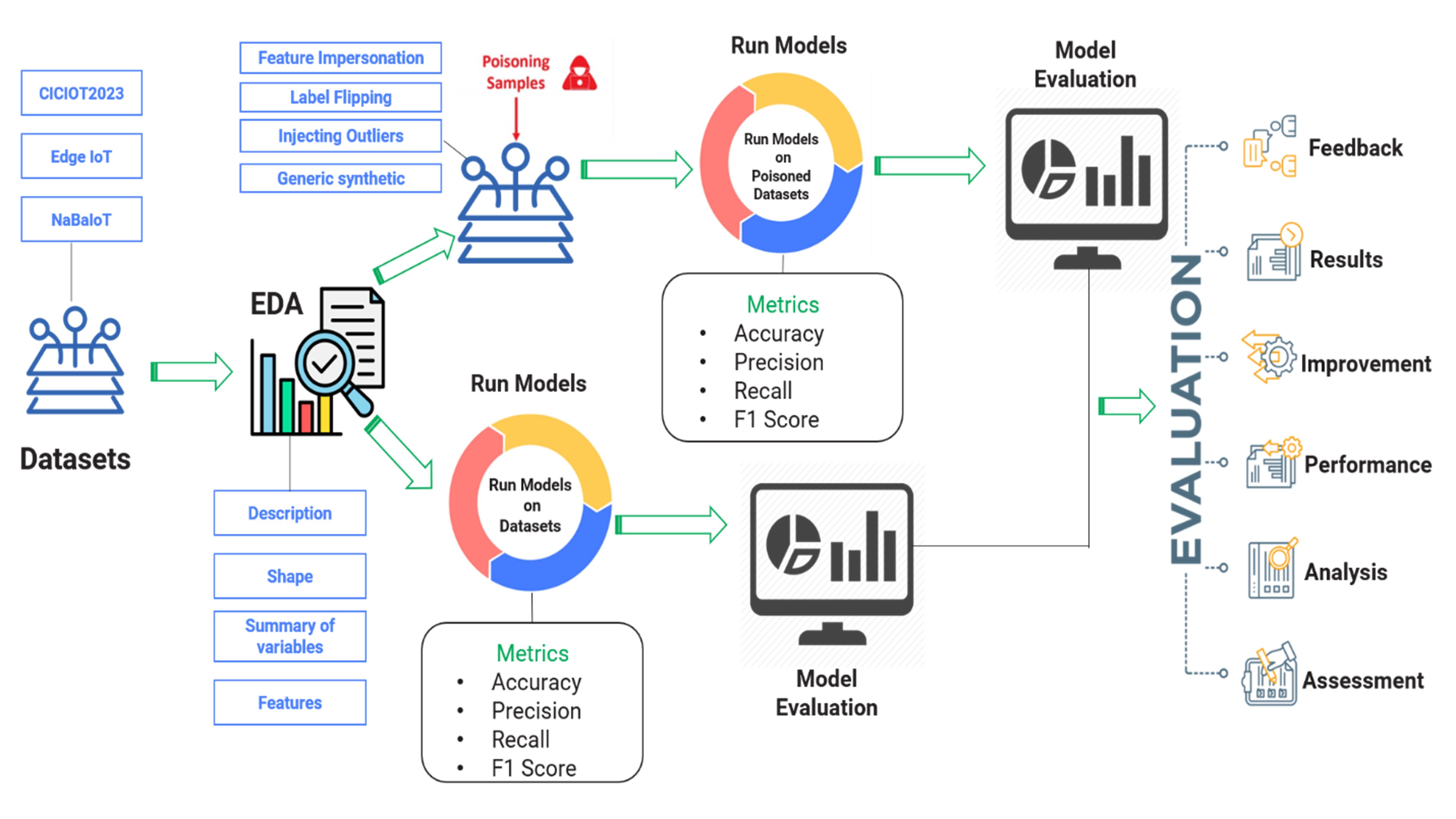}
    \caption{Overview of the Methodology for Evaluating ML-Based NIDS Under Data Poisoning Attacks}
    \label{fig3}
\end{figure*}

\sloppy
This section outlines the methodological framework adopted to evaluate the resilience of machine learning-based Network Intrusion Detection Systems (NIDS) against data poisoning attacks in Internet of Things (IoT) environments. Building on prior work in intrusion detection \citep{Kumar2022NIDSMachineLearning, Yaokumah2021IoTNIDS}, adversarial ML \citep{Biggio2012PoisoningSVM, Yuan2019AdversarialExamples}, and IoT cybersecurity datasets \citep{Ferrag2022EdgeIIoTset, Carlos2023CICIoT2023}, the methodology integrates dataset preprocessing, exploratory data analysis, supervised learning, adversarial manipulation, and performance evaluation. Figure~\ref{fig3} provides an overview of the workflow.

The overall objective is to understand how adversaries can tamper with training data to degrade model performance, distort decision boundaries, or introduce stealthy misclassifications.

\subsection{Datasets}
\label{subsec3.1}
High-quality and diverse datasets are essential for realistic intrusion detection research, particularly when evaluating adversarial manipulation. This study employs three widely used IoT cybersecurity datasets—CICIoT2023, Edge-IIoTset, and N-BaIoT—which collectively cover consumer IoT, industrial IoT, and botnet-driven attack scenarios \citep{Nazir2023IoTBotnetReview, Abbasi2021NBAIoTAnomaly, Almseidin2022CyberMultiStep}. CICIoT2023 provides over one million traffic records from 105 devices and 33 attack types, offering a rich set of multiclass intrusion patterns \citep{Carlos2023CICIoT2023}. Edge-IIoTset contributes 1176 network and system-level features spanning 14 IIoT attacks, making it suitable for evaluating poisoning in heterogeneous industrial environments \citep{Ferrag2022EdgeIIoTset, Chen2020FederatedIDS, Li2021DeepFed}. N-BaIoT includes more than seven million time-series samples generated from devices infected with Mirai and Bashlite botnets, enabling detailed analysis of poisoning effects on anomaly detection \citep{Abbasi2021NBAIoTAnomaly}. Together, these datasets provide a comprehensive foundation for assessing the impacts of poisoning across various IoT traffic landscapes.

\subsection{Exploratory Data Analysis}
\label{subsec3.2}
Exploratory Data Analysis (EDA) was conducted to understand the structure and quality of the IoT datasets prior to model training. This involved examining dataset dimensions, variable types, and class distributions, as well as identifying the most relevant features for intrusion detection \citep{Muthukrishnan2017OutlierSurvey}. Missing values, abnormal patterns, and natural outliers were assessed to ensure data consistency and to establish a baseline for comparing poisoning-induced anomalies. Standard preprocessing steps, including normalization and categorical encoding, were applied to prepare the data for supervised learning. As highlighted in prior work, thorough EDA is essential for building reliable intrusion detection models and for interpreting the impact of adversarial manipulation \citep{Sarcevic2022CybersecurityXAI, DashIoTBiomedicalReview}.

\subsection{Machine Learning Algorithms}
\label{subsec3.3}
Four supervised machine learning models were employed to evaluate how different architectures respond to data poisoning in IoT intrusion detection. Random Forest (RF) was used as a robust ensemble baseline, enabling us to assess how tree-based aggregation mitigates noisy or corrupted samples \citep{AnalyticsVidhya2021RandomForest}. Gradient Boosting Machines (GBM) were included to examine how sequential error-corrective learning behaves when exposed to subtle adversarial perturbations \citep{Sarcevic2022CybersecurityXAI}. Logistic Regression (LR), a lightweight model commonly used in edge IoT deployments, enabled us to analyze how linear classifiers degrade when poisoning shifts decision boundaries \citep{Abbasi2021NBAIoTAnomaly}. Finally, Deep Neural Networks (DNNs) were used to evaluate the sensitivity of high-capacity models to adversarial manipulation, reflecting their growing role in modern IDS solutions \citep{Yuan2019AdversarialExamples}.

\begin{figure}[!t]
    \centering
    \includegraphics[width=\columnwidth]{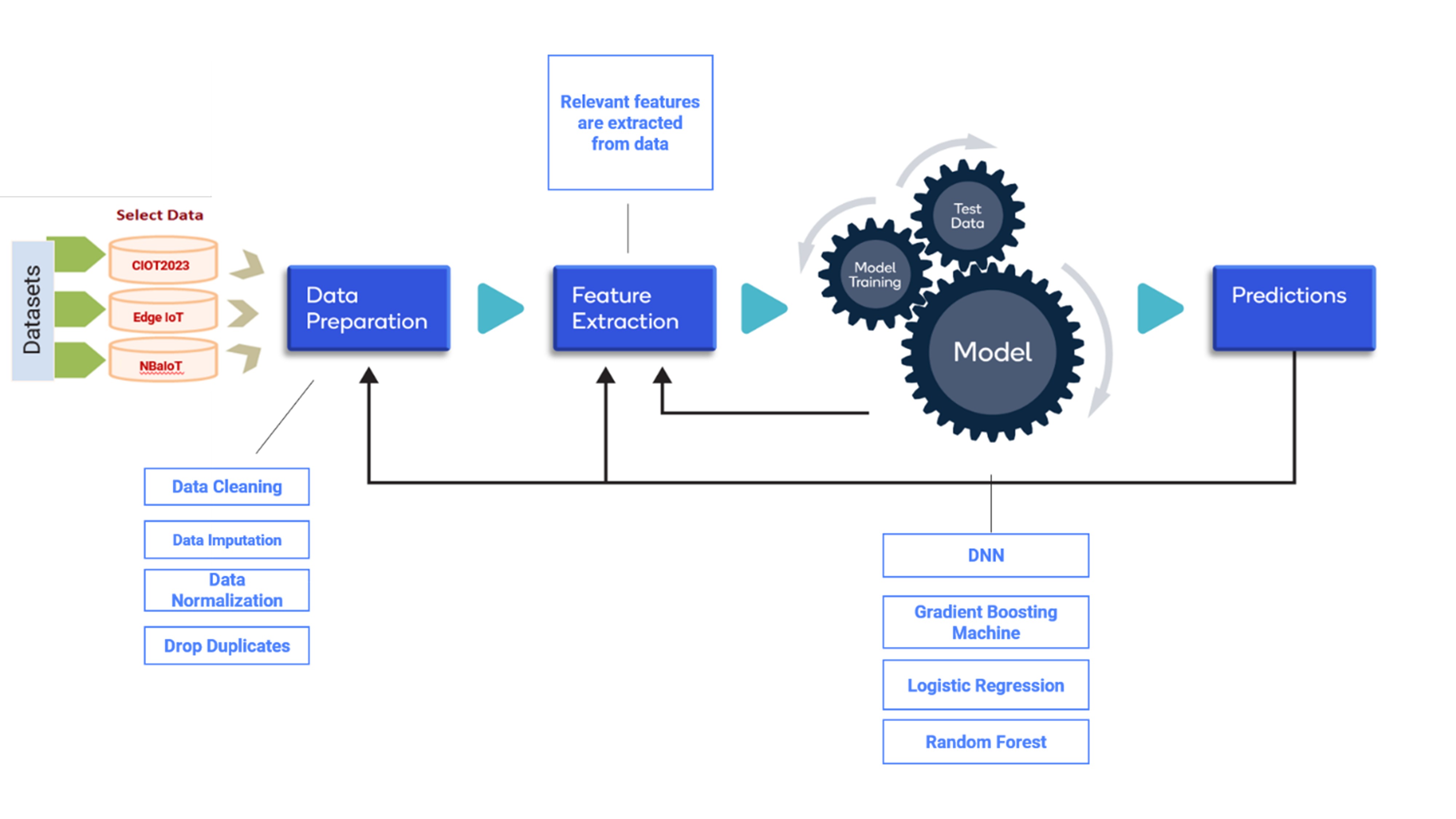}
    \caption{Machine Learning Model Development Pipeline}
    \label{fig5}
\end{figure}
These diverse models enable a comprehensive comparison of susceptibility to poisoning, ranging from lightweight, IoT-compatible classifiers to more advanced architectures.
\subsubsection{Experimental Environment}
\label{subsubsec3.3.2}

All experiments were conducted using Python on a Windows 11 workstation equipped with 16 GB RAM and an Intel Core i7 processor. Data preprocessing and EDA were performed using \texttt{pandas} and \texttt{matplotlib}, while ML models were implemented using the \texttt{scikit-learn} framework. Model configurations followed widely accepted hyperparameter settings from previous intrusion detection studies \citep{Sugi2020MLTechniquesIoTIDS, Dunn2020PoisoningRobustness}.

\subsubsection{Model Development Process}
\label{subsubsec3.3.3}
Figure~\ref{fig5} shows the pipeline for model development, including data ingestion, preprocessing, splitting, training, and evaluation. Techniques such as normalization, feature scaling, and class balancing enhance model performance and mitigate learning bias \citep{DashIoTBiomedicalReview}.

\subsection{Performance Metrics and Data Poisoning Techniques}
\label{subsec3.4}

\begin{table*}[t!]
\centering
\small
\caption{Comparison of Data Poisoning Techniques in ML-Based IoT Intrusion Detection}
\renewcommand{\arraystretch}{1.15}
\setlength{\tabcolsep}{4pt}

\resizebox{\textwidth}{!}{
\begin{tabular}{l|p{4.2cm}|p{4.2cm}|p{4.2cm}|p{4.2cm}}
\hline\hline
\textbf{Category} 
& \textbf{Feature Impersonation} 
& \textbf{Synthetic Outliers} 
& \textbf{Outlier Injection} 
& \textbf{Label Flipping} \\
\hline\hline

\textbf{Manipulation Method} 
& Alters selected feature values so malicious samples imitate benign traffic patterns 
& Introduces fabricated, statistically abnormal samples into the dataset 
& Injects extreme or noisy values to distort the feature distribution 
& Reassigns ground-truth labels to incorrect classes \\

\textbf{Impact on Data} 
& Blurs the separation between benign and malicious behavior 
& Shifts dataset distributions and increases variance 
& Creates instability and irregularity in feature space 
& Corrupts supervision and weakens class separability \\

\textbf{Objective} 
& Evade identification by mimicking normal behavior 
& Decrease prediction reliability and robustness 
& Trigger misclassification through noise-driven confusion 
& Manipulate learned boundaries and degrade detection accuracy \\

\textbf{Detection Difficulty} 
& High — subtle feature-level manipulation 
& Moderate — detectable with distribution checks 
& High — outliers may be masked within variability 
& Low–moderate — requires careful validation of labels \\
\hline\hline
\end{tabular}}
\label{tab:data_poisoning_comparison}
\end{table*}

In this work, model performance was evaluated using four widely adopted intrusion-detection metrics \citep{Nazir2023IoTBotnetReview, HodoANNIDSThreatAnalysis}. Accuracy measured the overall correctness of predictions, precision assessed how reliably attacks were identified, recall quantified the model’s ability to detect true malicious events, and the F1-score balanced both factors—supporting consistent comparisons of all models across clean and poisoned IoT datasets. To assess adversarial robustness, four poisoning techniques were applied following established practices in adversarial machine learning \citep{Biggio2012PoisoningSVM, Muthukrishnan2017OutlierSurvey, Zhang2021PoisonGAN}. These include: \textbf{Label Flipping}, which corrupts the ground-truth supervision; Outlier Injection, which introduces extreme or noisy samples; Feature Impersonation, where attackers modify feature values to mimic benign behavior; and Generic Synthetic Outliers, which distort dataset distributions using artificially generated samples. Collectively, these attacks represent diverse strategies for degrading model reliability, shifting decision boundaries, or masking malicious behavior. To provide a consolidated overview, Table~\ref{tab:data_poisoning_comparison} compares these poisoning strategies in terms of manipulation method, impact on data, adversarial objectives, and detection difficulty.


\subsection{ML Algorithms with Poisoned Data}
\label{subsec3.5}

After applying the poisoning strategies, each poisoned dataset was used to retrain RF, GBM, LR, and DNN models. The evaluation aims to quantify performance degradation and identify model-specific vulnerabilities, aligning with prior assessments of poisoning in cybersecurity \citep{Dunn2020PoisoningRobustness, Chiba2020PoisonDefense}. This analysis provides insight into how different models respond to varying levels of contamination, revealing susceptibility trends and highlighting the importance of robust training strategies. The findings inform the development of hybrid defensive mechanisms integrating adversarial training, anomaly-based filtering, and feature-level validation.


\section{Experimental Findings and Analysis}
\label{sec4}
This section presents the experimental findings, beginning with a review of EDA visualizations to contextualize the datasets, followed by an evaluation of baseline model performance on clean data. The results are then contrasted with outcomes under poisoned conditions, illustrating how adversarial manipulation affects classification behavior. Overall, the analysis highlights differing levels of model resilience and underscores the security risks data poisoning poses to ML-based NIDS.

\subsection{Exploratory Dataset Analysis}\label{subsec4.1}
A consolidated exploratory analysis was performed on the CIOT2023 \citep{Carlos2023CICIoT2023}, Edge-IIoTset \citep{Ferrag2022EdgeIIoTset}, and N-BaIoT \citep{Abbasi2021NBAIoTAnomaly} datasets to understand their structure and suitability for supervised intrusion detection. All three datasets were complete, with no missing values, and were dominated by numeric features—an advantageous property for feature-driven ML models.

\begin{figure}[!t]%
\centering
\includegraphics[width=\columnwidth]{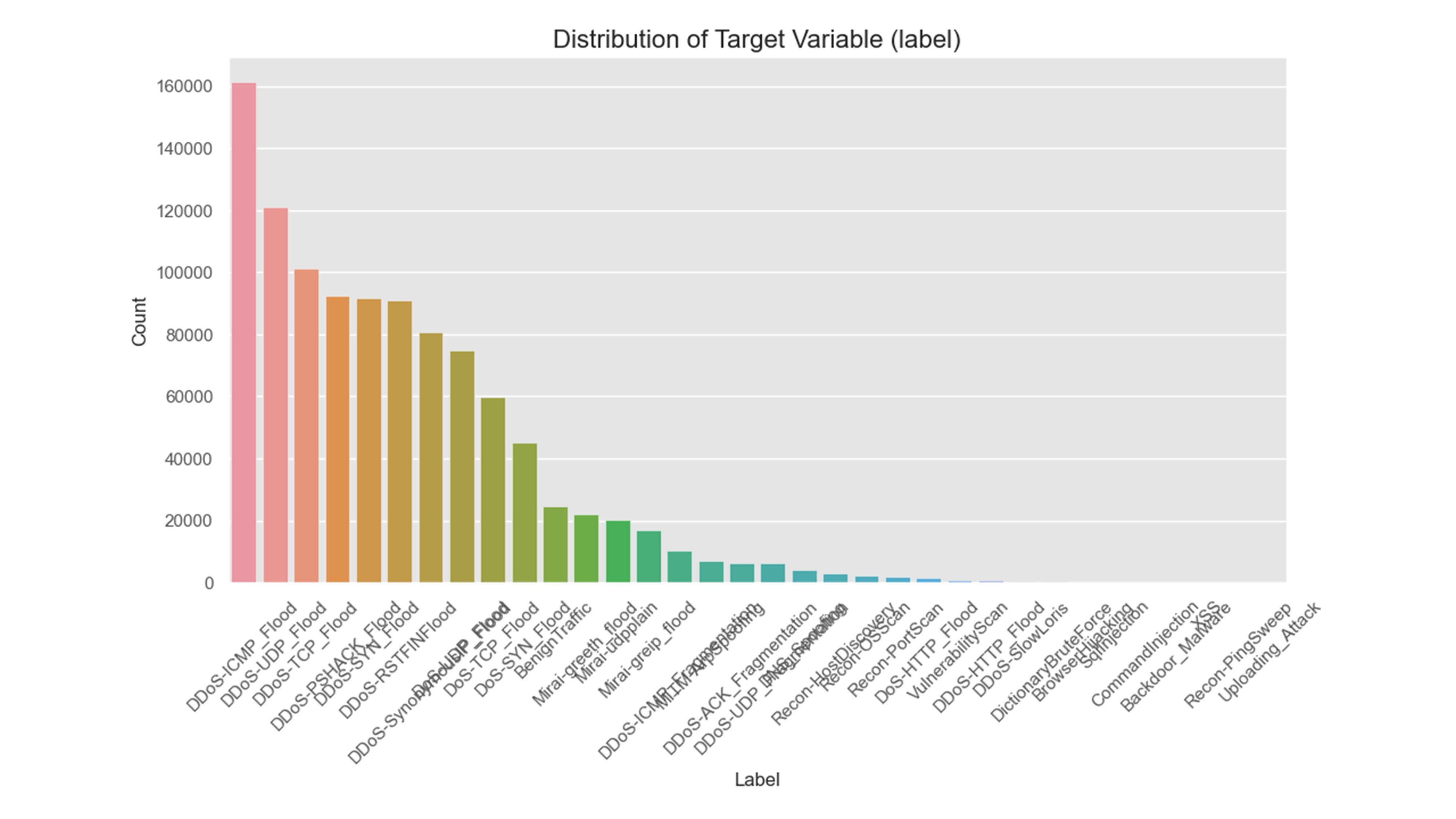}
\caption{Number of features per label - CICIoT2023}\label{fig7}
\end{figure}

\begin{figure}[!t]%
\centering
\includegraphics[width=\columnwidth]{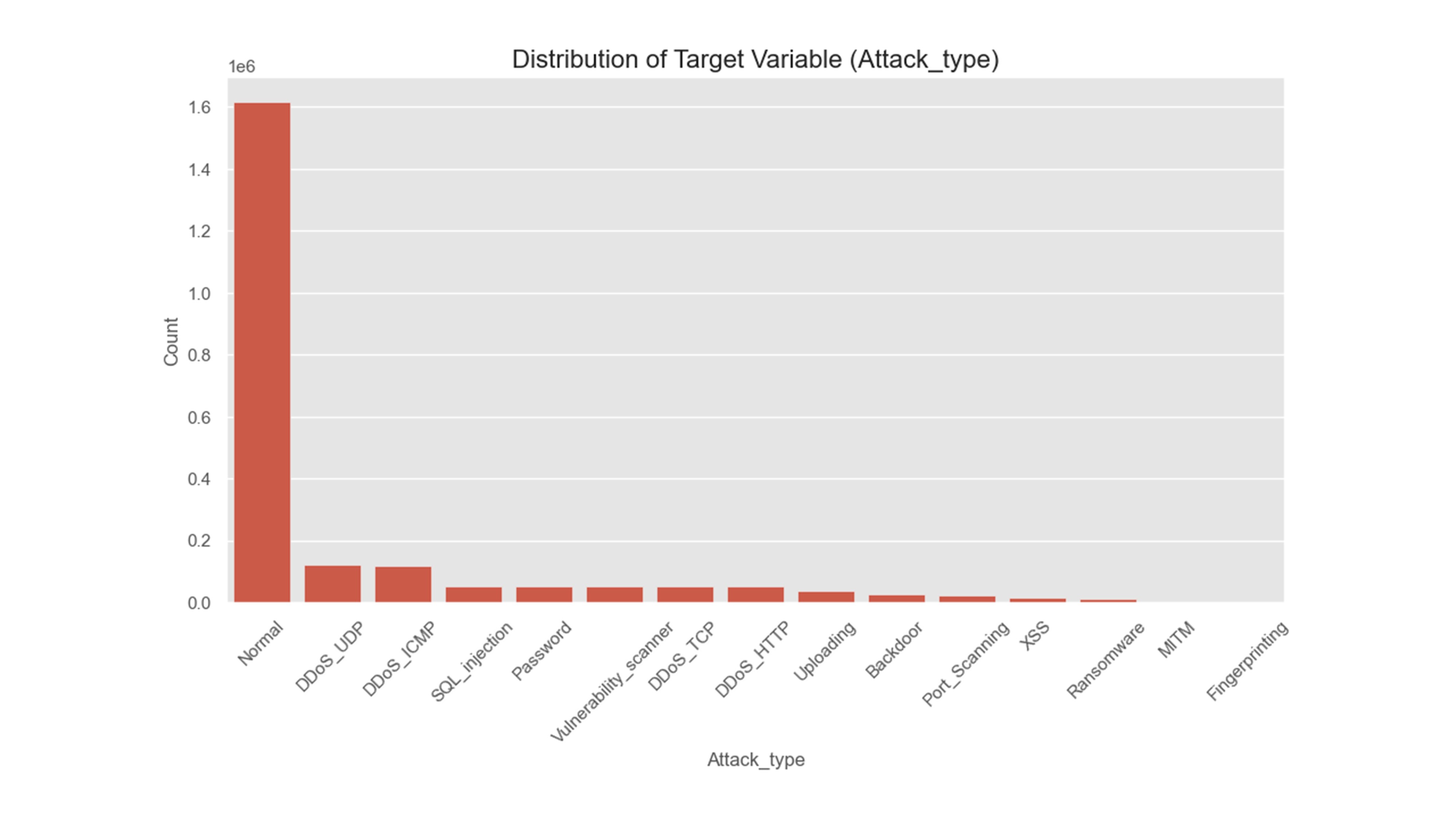}
\caption{Number of features per label – Edge IoT}\label{fig8}
\end{figure}

\begin{figure}[!t]%
\centering
\includegraphics[width=\columnwidth]{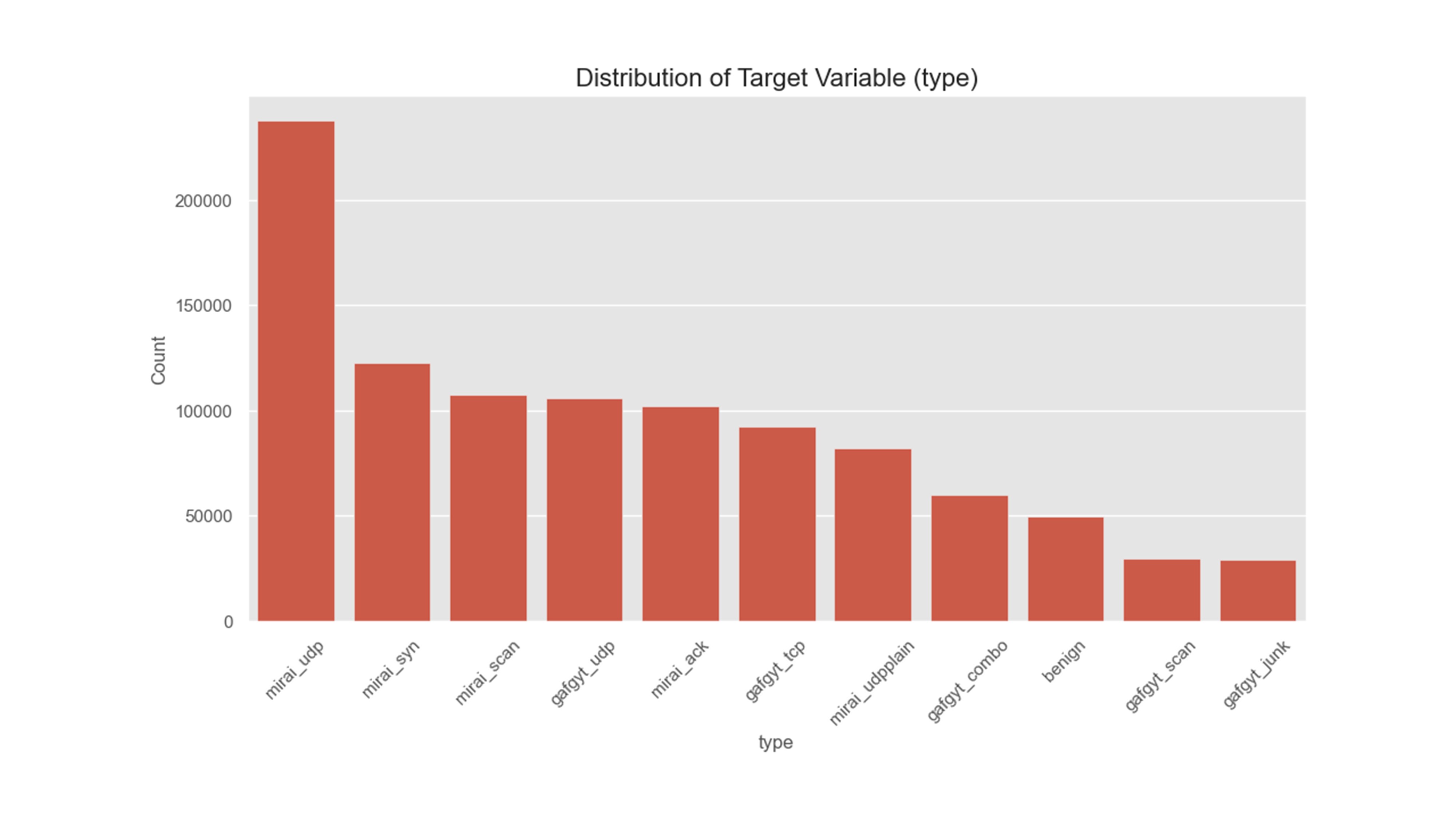}
\caption{Number of features per label – N-BaIoT}\label{fig9}
\end{figure}

CIOT2023 contains 1,048,575 instances and 47 features, with a single categorical target label. Its feature distribution across classes is illustrated in Figure~\ref{fig7}. Edge-IIoTset is larger and more heterogeneous, comprising 2,219,201 instances and 63 variables, including four categorical features; the corresponding class distribution is shown in Figure~\ref{fig8}. N-BaIoT offers the highest feature granularity, with 116 entirely numeric variables capturing device-level statistical behaviour, as depicted in Figure~\ref{fig9}.

Across all datasets, the absence of missing values and outliers, combined with high numeric dimensionality, indicates strong suitability for ML-based intrusion detection. CIOT2023 provides general IoT traffic, Edge-IIoTset expands industrial attack coverage, and N-BaIoT offers fine-grained telemetry. These distinctions support meaningful robustness comparisons under clean and poisoned training conditions. A consolidated summary appears in Table~\ref{tab:eda_summary}.

\begin{table*}[t!]
\centering
\small
\caption{Summary of Exploratory Data Analysis for IoT Datasets}
\renewcommand{\arraystretch}{1.5}
\setlength{\tabcolsep}{6pt}

\resizebox{\textwidth}{!}{
\begin{tabular}{l|p{2.5cm}|p{2.3cm}|p{2.3cm}|p{2.3cm}|p{2.0cm}|p{2.0cm}|p{2.0cm}}
\hline\hline
\textbf{Dataset} 
& \textbf{Total Instances} 
& \textbf{Total Variables} 
& \textbf{Missing Values} 
& \textbf{Numeric Columns} 
& \textbf{Categorical Columns} 
& \textbf{Target Variable} 
& \textbf{Outliers Detected} \\
\hline\hline

CIOT2023 
& 1,048,575 
& 47 
& None 
& 46 
& 1 
& \texttt{label} 
& None \\

Edge-IIoTset 
& 2,219,201 
& 63 
& None 
& 59 
& 4 
& \texttt{Attack\_label} 
& None \\

N-BaIoT 
& 1,018,298 
& 116 
& None 
& 116 
& 1 
& \texttt{type} 
& None \\
\hline\hline
\end{tabular}}
\label{tab:eda_summary}
\end{table*}

\subsection{Machine Learning Models Results on Datasets}
\label{subsec4.2}

\begin{figure*}[!t]
    \centering
    \makebox[\textwidth][c]{%
        \includegraphics[
            width=1.50\textwidth,
            height=0.56\textheight,
            keepaspectratio=false
        ]{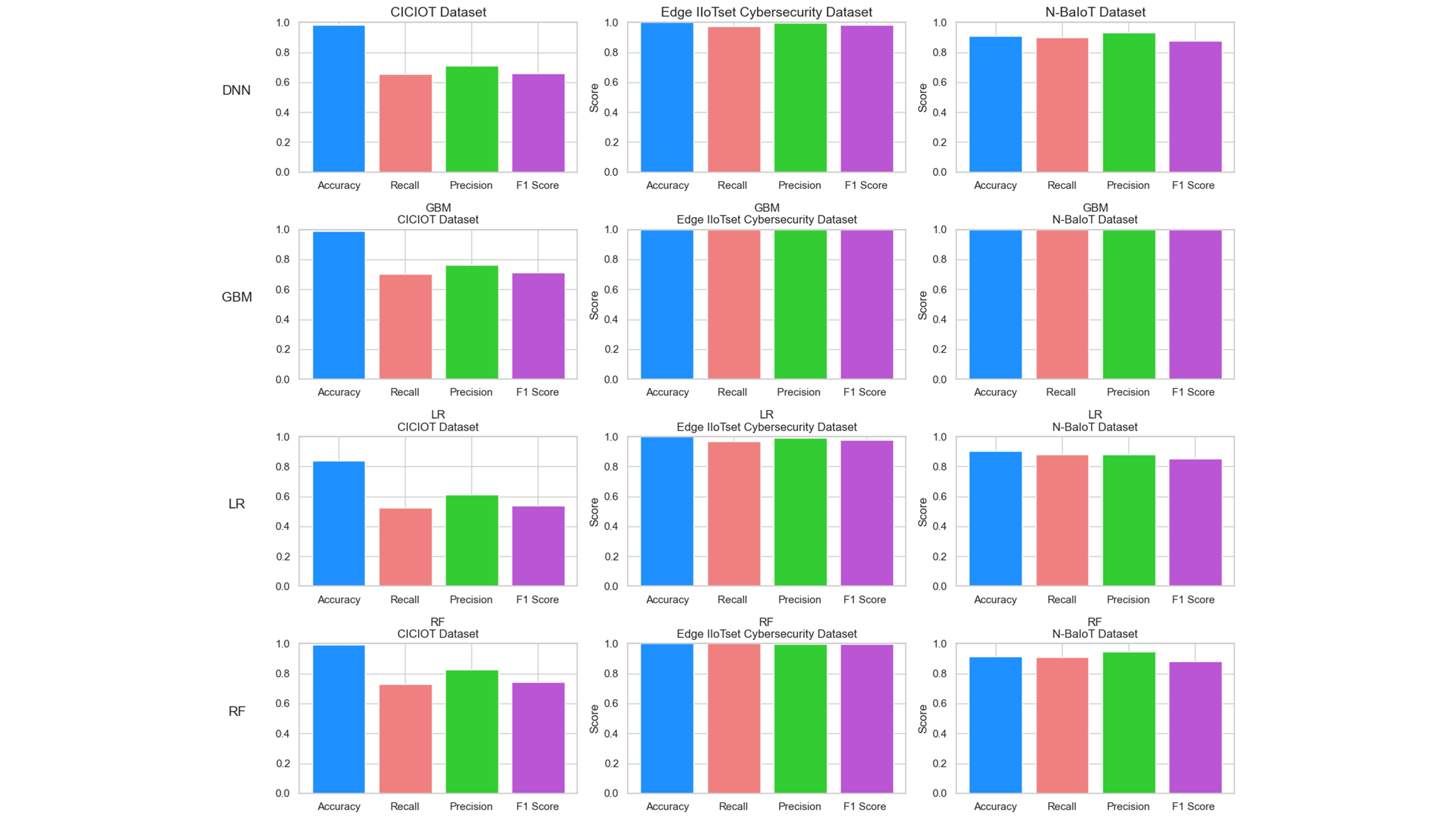}
    }
    \caption{Performance of ML Models on Original IoT Datasets}
    \label{fig10}
\end{figure*}

This subsection evaluates the performance of DNN, GBM, LR, and RF across the CIOT2023 \citep{Carlos2023CICIoT2023}, Edge-IIoTset \citep{Ferrag2022EdgeIIoTset}, and N-BaIoT \citep{Abbasi2021NBAIoTAnomaly} datasets using accuracy, precision, recall, and F1-score.

On CIOT2023, RF achieved the strongest performance with 99.29\% accuracy and a 74.24\% F1-score, followed by GBM at 99.06\% accuracy and a 71.06\% F1-score. LR performed considerably lower, with 83.88\% accuracy and a 53.89\% F1-score, indicating difficulty handling the dataset’s complex distributions.

Across the Edge-IIoTset dataset, all models—DNN, RF, and GBM—reached consistently high scores, with accuracy, precision, and recall near or above 99\%. This reflects the dataset’s clear class structure and rich feature space.

For N-BaIoT, GBM delivered exceptional performance with 99.96\% accuracy and a 99.91\% F1-score. RF also performed strongly, achieving 91.54\% accuracy and an 88.20\% F1-score. LR again lagged behind, with 90.30\% accuracy and an 85.35\% F1-score.

Overall, the comparison in Fig.~\ref{fig10} shows that RF and GBM consistently outperform DNN and LR across all datasets, demonstrating superior robustness and generalization in IoT intrusion detection tasks. LR remains the least resilient model, particularly in datasets with complex or high-dimensional features.


\subsection{Machine Learning Models Performance on Poisoned Datasets}\label{subsec4.3}

This section examines how four poisoning techniques—Outlier Injection, Label Flipping, Feature Impersonation, and Generic Synthetic Outliers—affect the predictive performance of GBM, LR, DNN, and RF across the CICIoT, Edge-IIoTset \citep{Ferrag2022EdgeIIoTset}, and N-BaIoT \citep{Abbasi2021NBAIoTAnomaly} datasets. By evaluating accuracy, precision, recall, and F1-score under adversarial contamination, the analysis highlights model vulnerabilities and resilience patterns.

\subsubsection{Robustness of Models on Label Flipping}\label{subsubsec4.3.1}

\begin{figure*}[!t]
    \centering
    \makebox[\textwidth][c]{%
        \includegraphics[
            width=1.50\textwidth,
            height=0.56\textheight,
            keepaspectratio=false
        ]{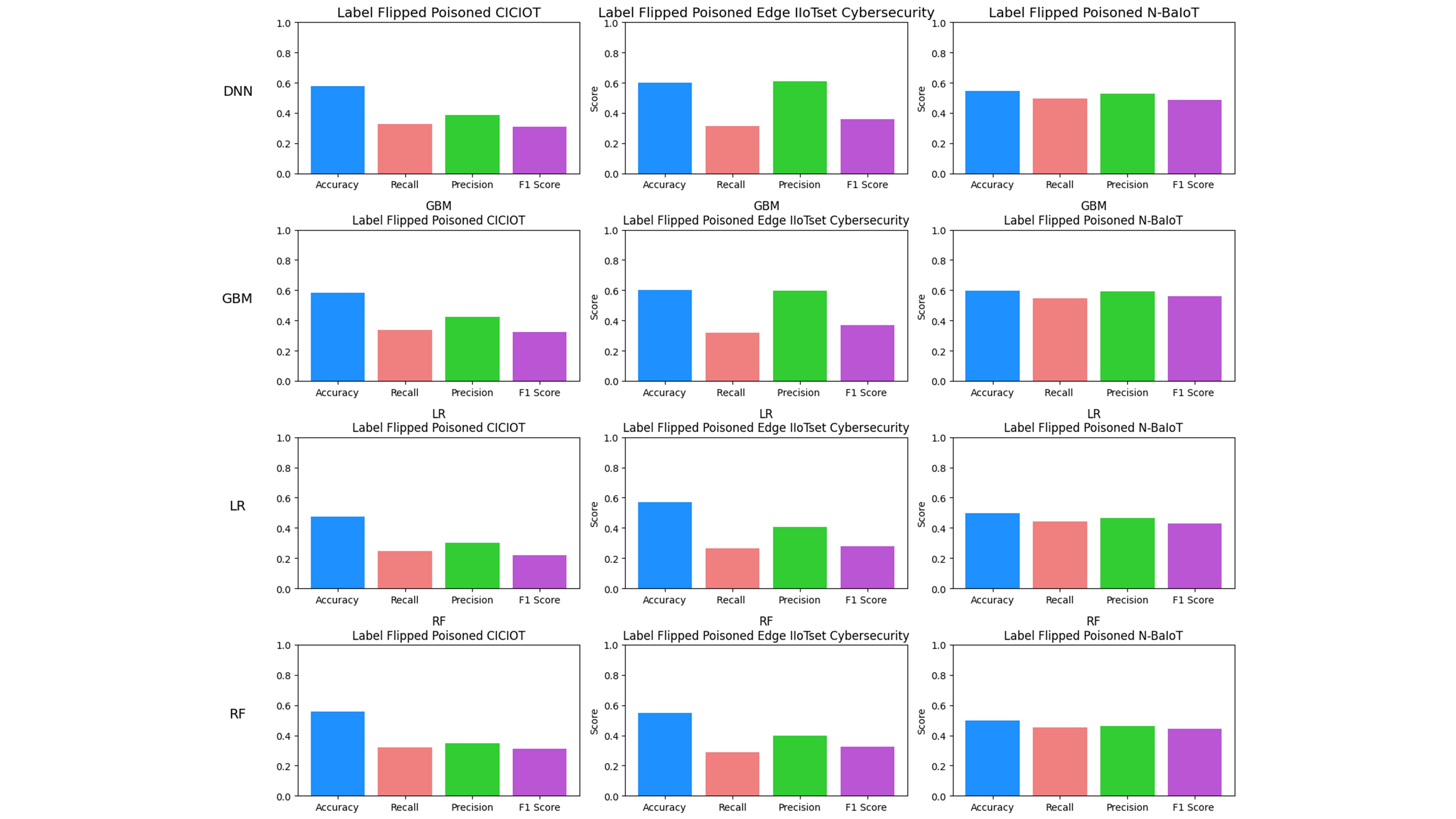}
    }
    \caption{Performance of ML Models on Label-Flipped Poisoned Datasets}
    \label{fig11}
\end{figure*}

Label flipping had the strongest negative impact across all datasets. As shown in Fig.~\ref{fig11}, accuracy dropped by up to 40\% in severe cases, with LR and DNN experiencing the largest declines. RF and GBM showed comparatively better stability but still recorded notable performance losses. These results highlight the disruptive nature of corrupted labels and underscore the importance of adversarial training and label verification strategies.

\subsubsection{Robustness of Models on Outlier Injection}\label{subsubsec4.3.2}

\begin{figure*}[!t]
    \centering
    \makebox[\textwidth][c]{%
        \includegraphics[
            width=1.50\textwidth,
            height=0.56\textheight,
            keepaspectratio=false
        ]{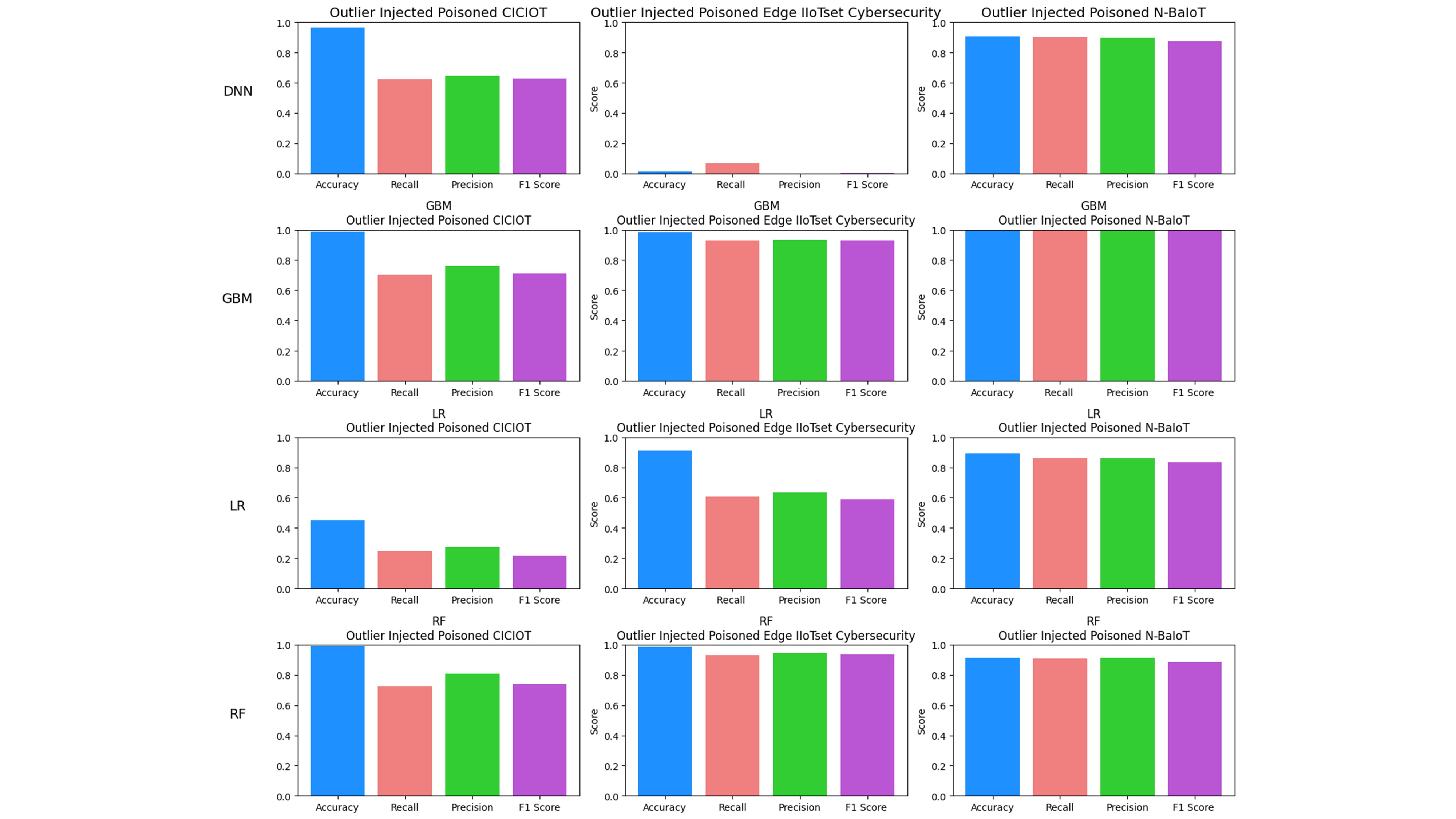}
    }
    \caption{Performance of ML Models on Outlier-Injected Poisoned Datasets}
    \label{fig12}
\end{figure*}

Outlier injection produced mixed effects depending on the dataset. As illustrated in Fig.~\ref{fig12}, RF and GBM preserved high performance in Edge-IIoTset and N-BaIoT, while LR and DNN suffered significant precision and recall degradation in CICIoT. The CICIOT dataset showed the largest vulnerability, with LR experiencing marked drops across all metrics. These findings highlight the importance of noise-aware preprocessing and feature-level sanitization.

\subsubsection{Robustness of Models on Feature Impersonation}\label{subsubsec4.3.3}

\begin{figure*}[!t]
    \centering
    \makebox[\textwidth][c]{%
        \includegraphics[
            width=1.50\textwidth,
            height=0.56\textheight,
            keepaspectratio=false
        ]{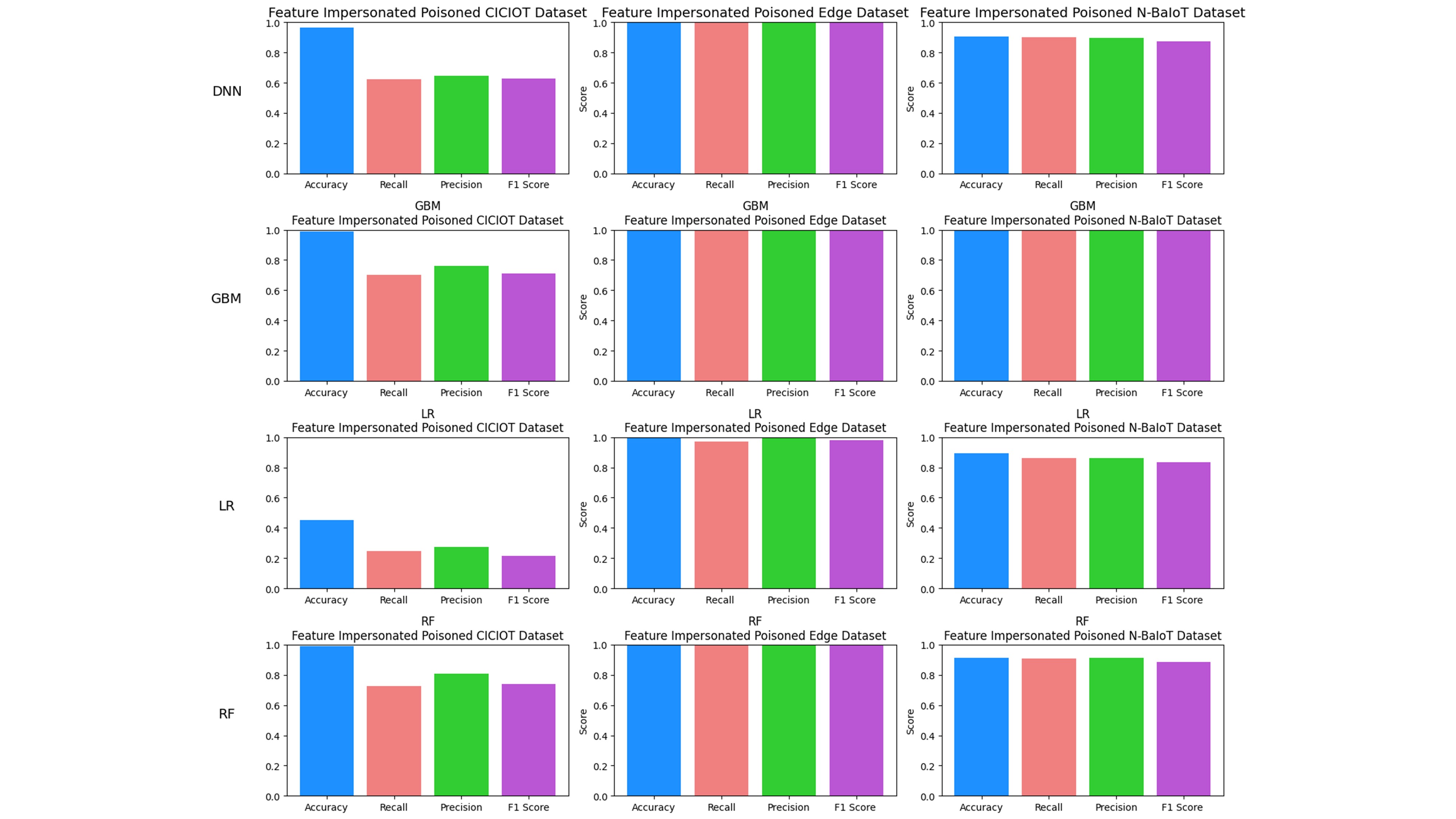}
    }
    \caption{Performance of ML Models on Feature-Impersonated Poisoned Datasets}
    \label{fig13}
\end{figure*}

Feature impersonation primarily affected models trained on CICIoT, where LR and DNN experienced sharp reductions in accuracy, recall, precision, and F1-score. As shown in Fig.~\ref{fig13}, RF and GBM were more resilient. In Edge-IIoTset and N-BaIoT, feature impersonation produced minimal disruption, indicating that its effectiveness depends on dataset complexity and feature dimensionality. These results underscore the need for feature-integrity checks in complex IoT environments.

\subsubsection{Robustness of Models on Generic Synthetic Outliers}\label{subsubsec4.3.4}

\begin{figure*}[!t]
    \centering
    \makebox[\textwidth][c]{%
        \includegraphics[
            width=1.50\textwidth,
            height=0.56\textheight,
            keepaspectratio=false
        ]{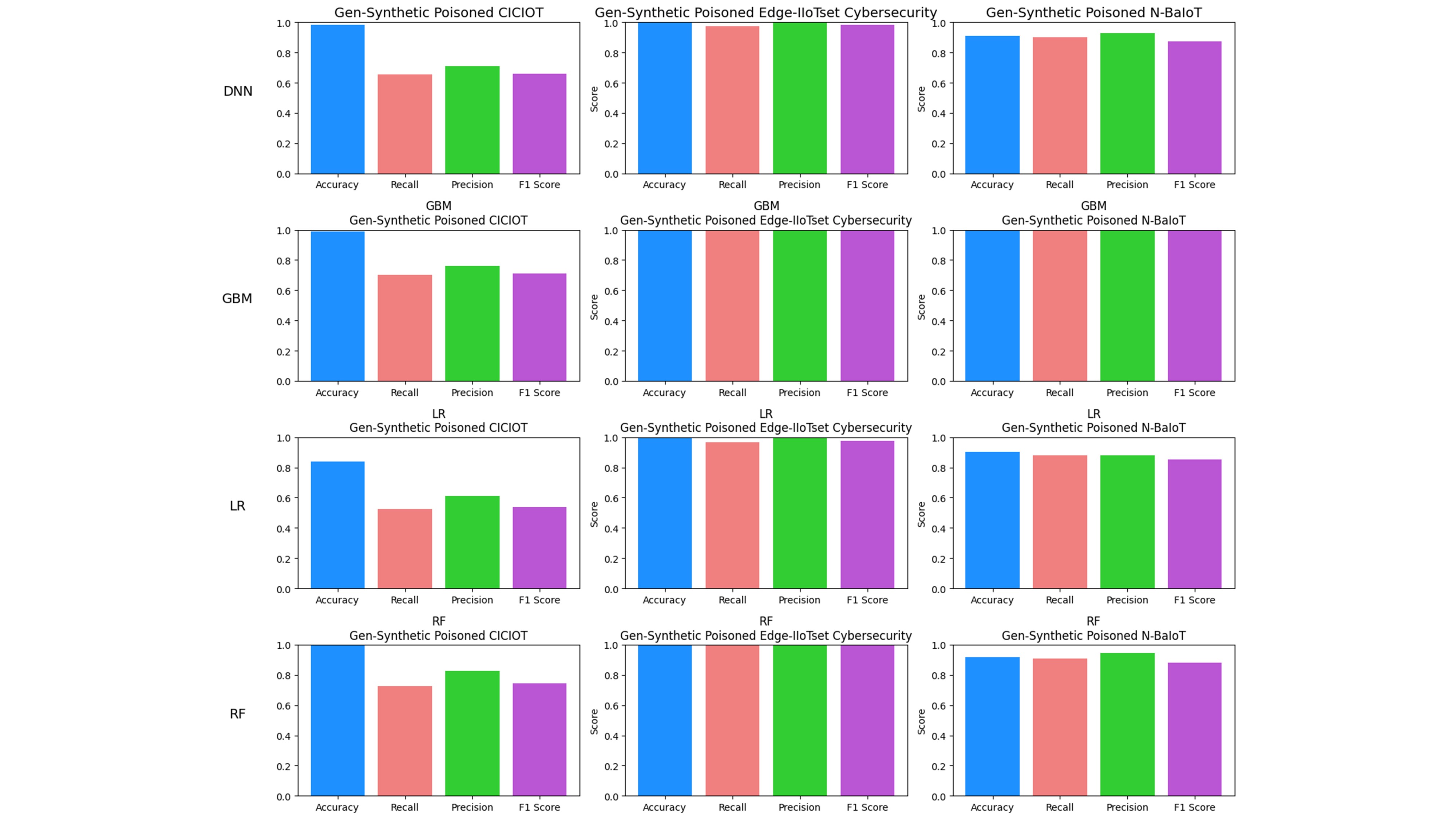}
    }
    \caption{Performance of ML Models on Generic Synthetic Outlier Poisoned Datasets}
    \label{fig14}
\end{figure*}

Synthetic outliers had milder effects than structured attacks. As shown in Fig.~\ref{fig14}, LR saw moderate degradation in CICIoT and slight declines in N-BaIoT, while GBM, RF, and DNN remained largely stable across datasets. Compared to label flipping, these perturbations caused less severe disruption, suggesting that models tolerate random noise better than targeted manipulation. This highlights the role of anomaly detection and robust preprocessing as lightweight safeguards.

\subsubsection{Summary of Models Robustness on Poisoned Datasets}\label{subsubsec4.3.5}

Overall, poisoning exposed clear differences in algorithmic resilience. Label flipping was the most damaging attack, severely degrading LR and DNN, especially on CICIoT. RF and GBM consistently proved more robust across all techniques and datasets. Outlier-based attacks affected models unevenly, with their severity depending on dataset complexity. No model demonstrated universal resistance, underscoring the necessity for layered defenses, feature-aware preprocessing, and ensemble-based approaches to secure ML-driven IoT intrusion detection systems.

\section{Conclusion}
\label{sec:conclusion}
This research investigated the vulnerability of machine learning models to data poisoning attacks in IoT intrusion detection, utilizing three major cybersecurity datasets: CICIoT, Edge-IIoT, and N-BaIoT. Our analysis revealed distinct resilience patterns: Gradient Boosting Machine and Random Forest consistently demonstrated strong robustness, especially against Label Flipping and Generic Synthetic Outlier attacks, whereas Logistic Regression and Deep Neural Networks were notably more susceptible. These results highlight the urgent need for adversarially resilient ML models in IoT security, reinforcing the broader importance of data integrity, real-time threat detection, and adversarial risk mitigation. The findings also carry meaningful implications for policymakers and cybersecurity practitioners, emphasizing the need for standardized evaluation frameworks and adaptive defense strategies that can preemptively detect and neutralize adversarial manipulation. Looking ahead, strengthening ML-based intrusion detection will require real-time adversarial detection mechanisms, adaptive learning strategies that help models recover after exposure to poisoned inputs, and hybrid approaches that combine ensemble learning with federated learning for decentralized and secure training. Equally essential are realistic, large-scale attack simulations on live IoT environments to validate model robustness outside controlled settings. Finally, advancing data integrity through techniques such as differential privacy and blockchain-based validation will be critical for ensuring long-term trust and resilience in IoT security ecosystems.














\bibliographystyle{unsrt}
\bibliography{ref}

\end{document}